# Report from the Workshop on
# Molecular Spectroscopy in the Era of Far-Infrared Astronomy

held at Emory University, Atlanta, GA
October 28 – 31, 2012

Report prepared by:
Susanna Widicus Weaver, Emory University
Joel Bowman, Emory University,
Michael Duncan, University of Georgia
Darek Lis, California Institute of Technology
John Pearson, NASA Jet Propulsion Laboratory
Steven Shipman, New College of Florida
Phillip Stancil, University of Georgia
Al Wootten, National Radio Astronomy Observatory

Conference Organizing Committee:
Susanna Widicus Weaver (Emory University) – Meeting Chair
Joel Bowman (Emory University)
Michael Duncan (University of Georgia)
Thomas Orlando (Georgia Institute of Technology)
Steven Shipman (New College of Florida)
Phillip Stancil (University of Georgia)
David Schultz (University of North Texas)
Glenn Wahlgren (NASA Headquarters)
Al Wootten (National Radio Astronomy Observatory)

# I. Executive Summary

The development of the next generation of far-infrared astronomical instrumentation has placed molecular astrophysics research at a crossroads. The Herschel, SOFIA, and ALMA observatories are providing spectral observations that have sensitivity limits well below those of any preceding ground- or space-based observations. These observatories are also operating at frequencies that are ideal for probing the molecular universe. The broadband spectral acquisition capabilities of these observatories will eliminate the tedious, one-line-at-a-time approach to molecular observations that has historically hampered ground-based radioastronomical identification of molecules. The ALMA observations will advance this field even further, providing highly spatially-resolved molecular information and imaging capabilities in addition to broadband spectral acquisition.

The expected amount of information from these new facilities is quite daunting because the laboratory spectral database that will enable interpretation of these observations is severely lacking. Likewise, the tools needed to analyze both the laboratory and observational spectra do not meet modern needs. Laboratory spectral surveys across the frequency ranges covered by these instruments have not been completed for even the most abundant interstellar molecules, commonly referred to as ``interstellar weeds,'' let alone the majority of the ~170 interstellar molecules identified to date, or new molecules that have not yet been identified. This lack of laboratory spectral information will limit interpretation of the observational spectra, which will be near the line-confusion limit, therefore limiting the scientific return from these observatories.

Fortunately, the hardware and technology advancements associated with the construction of these observatories have also spurred new developments in laboratory spectroscopic techniques. We therefore assembled at Emory University in October, 2012, to discuss the state-of-the art in molecular laboratory spectroscopic techniques used to study molecules of astrophysical interest, and highlight recent successes in laboratory efforts that complement new astronomical observations. The format of the meeting facilitated detailed discussion of the needs for collecting and interpreting molecular spectroscopic information to support observational astrophysics in the era of Far-IR astronomy. We present the findings and recommendations from this workshop in this report.

The participants in this workshop identified the development of new laboratory capabilities that offer rapid, broadband, high-resolution, high-sensitivity measurements of molecular spectra in the frequency ranges that overlap with new observational facilities as the top priority for far-infrared laboratory astrophysics initiatives over the next 5-10 years. An additional major need is the development of suitable analysis tools for broadband molecular spectra from both the laboratory and space; though such efforts should be conducted within the framework of larger collaborative funding structures, rather than through single-PI efforts, so as to provide tools that are generally applicable rather than highly specialized. Additional efforts in theory, computation, and modeling that support these goals are also important for advancing this field.

## II. Overview of the challenges faced from Far-IR astronomy

### 1. Lessons learned from Herschel

The Herschel Space Observatory,[1] launched on May 14, 2009, is the first space facility to fully cover the submillimeter spectral range from 60 to 670 µm, most of which is completely blocked from the ground by Earth's atmosphere. The high-resolution heterodyne spectrometer, HIFI, has provided velocity-resolved spectra of Galactic and extragalactic sources in the frequency range 480–1250 GHz and 1410–1910 GHz, divided into 7 mixer bands. With a frequency resolution of approximately 1.1 MHz over the full intermediate frequency band (4 GHz for the SIS mixer bands and 2.4 GHz for the two highest-frequency HEB bands), this corresponds to over a million independent spectral channels. One of the main HIFI science themes is "unbiased spectral line surveys", which contain the complete chemical inventory (for polar species), the chemical history and evolutionary state, the excitation and cooling conditions, as well as the dynamical picture of the objects surveyed. Submillimeter wavelengths give access to high-energy transitions, excited only in the immediate vicinity of newly formed stars, probing the full extent of the grain surface chemistry in regions where fresh ices are evaporating from grain mantles and can be studied through their rotational spectra. Two Herschel guaranteed time key programs focus specifically on spectral line surveys. HEXOS [Bergin et al. 2010] includes spectral scans of five sources in the Orion and Sagittarius B2 regions, while CHESS [Ceccarelli et al. 2010] includes observations of two prestellar cores, an outflow interaction region, a solar mass protostar, an intermediate mass protostar, and two high-mass sources. Together, these data sets cover a wide range of protostellar masses, luminosities, and evolutionary stages. Orion KL and Sgr B2(N) are the most line-rich sources on the sky, providing templates for other objects and ground-based follow-up studies. Observations of five sources within the Orion molecular cloud allow a direct comparison of the chemistry of different stages of embedded star formation within the same complex, under the same initial conditions. In addition, a complete HIFI spectral scan of the evolved star IRC+10216 has been carried out. The high spectral resolution ($\lambda/\Delta\lambda > 10^6$) allows disentangling multiple physical components with the HIFI beam, e.g. the hot core, quiescent, and shocked gas on the line of sight toward Orion KL. Observations of Sgr B2 allow studies of star formation, and chemistry, in an extreme environment, characterized by high density, gas pressure, turbulence tidal forces, strong UV, cosmic ray and X-ray fluxes. Fundamental rotational transitions of light hydrides and deuterides have high critical densities and are best studied in absorption toward bright submillimeter continuum sources. Several new hydrides have in fact been detected in the HIFI surveys (e.g. $OH^+$, $H_2O^+$, $H_2Cl^+$, $HCl^+$, ND). Some perplexing unidentified absorption lines have also been observed. A specific example is the strong absorption feature at 617 GHz that has components from all spiral arm clouds between Earth and the Galactic Center, but not the Sgr B2 envelope [Schilke et al. in prep.]. This feature does not match any known lines from previously identified interstellar molecules.

---

[1] Herschel is an ESA space observatory with science instruments provided by European-led Principal Investigator consortia and with important participation from NASA.

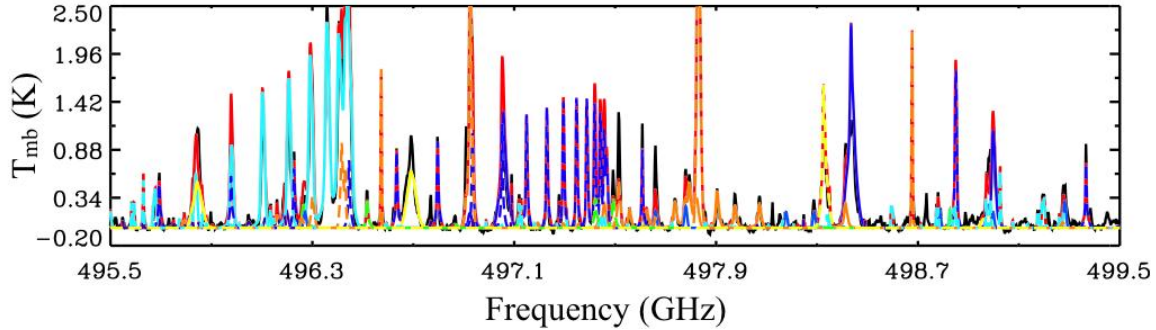

Figure 1— A 4 GHz-wide section of the Orion KL spectral scan (black) with the LTE fit superposed (color). From Crockett et al. [in prep.].

The immense HIFI datasets represent new modeling challenges. The classical line-by-line analysis approach, used in earlier ground-based studies, is simply no longer feasible. For example, the spectral scan of Orion KL contains approximately 20,000 spectral features, originating from different physical components within the HIFI beam (hot core, compact ridge, plateau). The lines are often blended and the complete spectrum has to be modeled at once (see Figure 1). However, such an analysis requires knowledge of precise line frequencies and strengths from laboratory measurements, which were largely unavailable for the HIFI frequency range just a few years before the beginning of the Herschel mission.

To address this challenge, a workshop on laboratory astrophysics in support of the Herschel Space Observatory was held in October 2006 at the California Institute of Technology in Pasadena, California. The final report[2] from this workshop detailed the importance of performing laboratory spectroscopy at frequency commensurate with Herschel, i.e. up to 1.9 THz for the abundant molecules with rich spectra. The most important four molecules (Class 1 "weeds") identified are $CH_3OH$ (methanol), $HCOOCH_3$ (methyl formate), $CH_3OCH_3$ (dimethyl ether), $CH_3CH_2CN$ (ethyl cyanide), and their isotopologues. Five somewhat less prominent molecules (Class 2 "weeds") are $C_2H_3CN$ (vinyl cyanide), $SO_2$ (sulfur dioxide), $CH_3CN$ (methyl cyanide), $HC_3N$ (cyanoacetylene), and $CH_3CHO$ (acetaldehyde).

Subsequently there was a Herschel call for laboratory astrophysics proposals, along with several rounds of the annual laboratory portion of the NASA APRA program. Similar opportunities also occurred in France, Germany, Canada, and Spain. A number of proposals were funded and some rather limited funding to extend laboratory frequency coverage was made available. The spectroscopic community made unprecedented efforts to collaborate and address this challenge for Herschel. The current state of the analysis for the aforementioned "weeds" is given in

---

[2] "Report from the Workshop on Laboratory Spectroscopy in Support of Herschel, SOFIA, and ALMA," http://www.submm.caltech.edu/labspec/Workshop%20Report%20Final.pdf

Appendix 1. To summarize these findings, each of these molecules has been characterized to the level required for interpretation of datasets from single dish observations such as those of Herschel and SOFIA. The one notable exception is an exploration of the intensities in the three fold internal rotation case where theory has never been experimentally tested. The result of this combined effort to complete the spectral analysis for these key interstellar molecules is one of the great successes of the Herschel program, and these results will be extremely important first steps in interpretation of ALMA and SOFIA spectra.

In additional to spectroscopic data provided by laboratory astrophysics, the analysis of the HIFI spectra requires modeling tools that allow computing simulated spectra of astronomical objects for a given set of physical conditions (source sizes, temperatures, column densities). Currently this is largely limited to local thermodynamic equilibrium (LTE), using dedicated software, while still allowing multiple spatial components within the beam. This approximation works quite well for line identification purposes in molecular hot cores (Figure 1), as only about 10% of mostly weak lines remain unidentified in the Herschel spectral scan of Orion KL.

Early on, some of the most prolific molecular line emitters, the interstellar "weeds" listed above, were often simply seen as an obstacle to identifying spectral features of "flowers"— species of particular astrophysical or astrochemical interest. However, this distinction between "weeds" and "flowers" seems no longer valid. With hundreds of emission lines covering a wide range of excitation conditions, the "weeds" can provide invaluable information about physical conditions in star-forming regions. One example is the analysis of the methanol band emission in Orion KL [Wang et al. 2011], which demonstrated that the Compact Ridge component is externally heated. HIFI data have also shown that species such as methyl and ethyl cyanide are particularly good probes of the hottest gas in the vicinity of newly formed stars.

Perhaps the most important lesson from Herschel/HIFI is that it is imperative to understand the needs early on and set specific goals that are achievable within the limited resources. Looking into the future, two broad goals can be identified. The first one is of immediate interest to the astrochemistry community. To fully understand chemistry of the interstellar medium, we need laboratory spectra of all the "flowers". These include some key ions, but also isomers and isotopologues of the abundant gas-phase species. Studies of isotopic ratios provide key information about the formation and evolution of ISM materials and their connection to the solar system. Such studies will be of particular interest for ALMA with its unparalleled sensitivity. Moreover, the laboratory data have to be easily accessible by the astronomical community and in a form that is easy to use. This in turn requires close collaborations between the two communities; with the laboratory spectroscopists working hand-in-hand with other members of the research teams analyzing astronomical spectra.

The second goal, with a broader appeal to the general astronomical community, has to do with direct astrophysical applications. It is increasingly clear that molecular spectra are fantastic tools to determine the physical conditions in the interstellar environments: the temperature, density, ionization fraction, etc. However, to achieve this goal, we need a good understanding of which species best trace different interstellar environments. One example is the use of deuterated isotopologues as tracers of the cold phases at the onset of star formation (prestellar cores). Moreover, to use molecules as quantitative tools, we need collisional rates, as well as 3D radiative transfer models for gas and dust. Good progress toward these goals is already being made. However, challenges do remain. For example, computations of collisional cross-sections are time and manpower intensive. A careful assessment of the priorities is thus needed. Finally, all these tools have to be easily accessible and incorporated into user-friendly software packages.

## 2. *What new challenges do we face with SOFIA and ALMA?*

While the broadband and sensitive Herschel datasets have presented a challenge to the spectroscopy community, the next generation of telescopes — ALMA, EVLA, GBT, SOFIA, and others — will produce even larger datasets at high spatial and spectral resolution over broad bandwidths. A recent estimate for ALMA suggests that for rather typical spatial and frequency resolution for a full 8 hour track, a dataset of 30 TB might result [Lacy & Halstead 2012]. A data cube from such a single pointing observation could measure $10^3 \times 10^3 \times 10^4$, making display, examination, and comparison of the spectral lines challenging. Visualization of datasets of this size challenges tools currently available for astronomy. Ideally one could superpose images of different spectral lines for various kinds of chemical analyses as well as for excitation analyses.

Already ALMA has produced an image cube of fairly low sensitivity covering about 30 GHz that contains images of the Orion region in some 4000 distinct spectral lines. Fortman et al. [2012] showed how features of the spectrum at a single position in this cube might be fit with a straightforward excitation model of ethyl cyanide ($CH_3CH_2CN$). Such studies require good laboratory spectral data on the molecule and, ideally, on its collisional excitation. The ALMA datacube contains thousands of spatial resolution elements. Ideally, interconnected models would lead to an image of physical conditions across the region with kinematic structures providing insight into its three dimensional structure.

Both visualization and modeling tools are needed for interpretative study of the data being produced by current and planned instruments. Variables would be adjusted to fit multidimensional physicochemical models to the multiaxial data within the astrophysical context of the source under investigation. Coupled with these needs for new analysis tools is the need for more complete spectral datasets that cover the entire frequency range accessed by these new instruments. As was demonstrated by Herschel/HIFI, these spectral datasets are needed for a variety of molecules so that the chemical and astrophysical information from both the "weeds" and the "flowers" can be used to optimize the science return from these new facilities.

**III. Summary of the findings from this meeting and major conclusions**

Building on the lessons learned from Herschel and the results anticipated from ALMA and SOFIA, several themes emerged from the presentations and the discussion at this meeting. The large datasets associated with broadband astronomical observations as well as broadband laboratory spectra require new analysis tools to be developed. In addition, new advances in astrochemistry and interpretation of far-infrared astronomical measurements require limits to be pushed beyond current capabilities in all associated areas of laboratory spectroscopy, computation, theory, and modeling. Beyond the new endeavors that must be undertaken to advance research in this field, the participants of the workshop recognized a tension that often arises between making datasets publicly available in the timescales they are needed to support astronomy, and the time lag that is often involved in finishing a complete analysis. Overall, the conclusion was that collaboration and synergistic efforts are crucial for tackling these research challenges. We expand upon the details for each of these themes below.

*1. Large datasets require new tools to be developed*

Broadband heterodyne receivers coupled with new spectrometer technology make it straightforward to acquire 4 – 12 GHz of high-resolution spectral data in one observational acquisition from both ground- and space-based observing platforms. This bandwidth essentially turns each new spectral line observation into a line survey, and enables acquisition of extremely broadband spectral coverage in a very short amount of observing time. Coupled with these advances in observational technology is a great leap forward in laboratory capabilities. It has become routine for a given spectral dataset in the far-infrared to cover hundreds of GHz in spectral bandwidth at high resolution, revealing tens-of-thousands of individual spectral lines. Several laboratory astrophysics research groups now have spectrometer capabilities that cover the entire millimeter/submillimeter band up to 1 THz (or higher, in a few cases). New broadband spectrometers are also opening up access to higher frequencies or larger spectral coverage, combined with rapid data acquisition capabilities.

Based on these technological advances, the field is facing an unprecedented challenge of spectral assignment for datasets that cover tens to hundreds of GHz in bandwidth. Development of new analysis tools is therefore a high priority. Aspects of new tool development that would be particularly helpful for both lab and observational research include automated spectral assignment/fitting routines and multi-dimensional visualization software. Coupled with these efforts are extreme needs for increased computational power and data storage capabilities, user-friendly interfaces that are accessible by more than one community (i.e. lab spectroscopy, theory, and observational astronomy), and ongoing maintenance and updating of spectral archives and databases.

While there is consensus that these new tools are critical needs for ensuring the science return from new observational facilities, development of such tools should not be relegated to a single-PI funding structure, nor should it necessarily be directly tied to funding for lab-related efforts. However, any large-scale collaborative funding program that focuses on laboratory astrophysics or far-infrared astronomy should include dedicated resources for the development of new analysis tools in addition to the new science being proposed. Ideally such efforts would directly involve the experts who will use the tools, i.e. spectroscopists and astronomers, in addition to those who are best-suited to develop these tools, i.e. computer scientists, programmers, and software engineers, and the tools developed should be generally applicable to a broad range of research problems.

*2. New advances in astrochemistry and far-IR measurements require limits to be pushed beyond current capabilities*

As stated in the executive summary, molecular astrophysics research is standing at a crossroads, as far-IR observational facilities now surpass most laboratory techniques in data acquisition rate and sensitivity. Laboratory techniques that offer rapid, broadband, high-resolution, high-sensitivity measurements of molecular spectra in the frequency ranges that overlap with new observational facilities are crucial for the advancement of this field. Such techniques are important for characterizing a large number of important astrophysical molecules, including excited vibrational and electronic states, unstable molecules that drive the chemistry of the ISM (such as radicals and ions), isotopologues, etc. However, only a handful of groups are working on new technique development. Such efforts should therefore be the top priority for funding in laboratory astrophysics over the next 5-10 years.

Going beyond the laboratory spectroscopic needs, several other areas of research that have overlap with both the lab and observing efforts also require great advancements in capabilities. Theoretical spectroscopy and modeling capabilities need to be extended to larger and floppier molecular systems. More realistic physical models of interstellar environments need to be coupled with improved reaction networks to more accurately model the chemistry of star-formation. Collisional information and reaction dynamics information are crucial for data interpretation and modeling, yet this information is incomplete even for the simplest of molecular systems. All of these studies require full-dimensional potential energy surfaces from theory, with refinements from experiment, and benchmarking dynamical measurements.

*3. Timely release of data, collaboration, and synergy are crucial for tackling the challenges of far-IR laboratory astrophysics.*

This field of research involves a small number of researchers, but their efforts are absolutely essential for supporting observational astronomy. The field must maintain its current size or grow in order to be sufficiently agile, so that the needs of far-infrared astronomy can be addressed. In order for any given researcher to succeed professionally, they must receive proper credit for their work, and this work must be published in a timely manner. However, the detailed and comprehensive nature of broadband datasets makes the timely release of data a significant challenge. Furthermore, making the information publicly available while ensuring that credit goes to the original authors is also a challenge in this field.

It is generally recognized that there is a tension between making datasets from laboratory astrophysics efforts publicly available in the timescales they are needed to support astronomy, and the time lag that is often involved in finishing a complete analysis. The question was raised at this meeting as to how "complete" an analysis must be to be useful. The answer to this question remains unclear at the present time, simply because of the limited number of cases that are available where a full analysis can be compared to an observational dataset. Therefore, the temptation in this field of research is to hold private databases of reaction information and spectra. The challenge here is that these databases cannot be vetted and have not always necessarily been peer-reviewed, yet are likely to be the most frequently used databases for observational data interpretation. Some efforts have been made to streamline existing databases; the Splatalogue effort spearheaded by the NRAO is one example, and some laboratory groups have indeed uploaded laboratory datasets to this database. However, release of the calibrated but unassigned laboratory datasets, rather than a complete analysis, brings with it an inherent risk in data interpretation. All datasets have flaws and idiosyncrasies that are often not easily relayed to people outside of the group that collected that information. In addition to these challenges, the researchers who collected the information included in public databases often do not receive proper credit for their work when these databases are used. One suggestion made at the workshop was that groups deposit and archive large datasets with journals, but again, this presents a challenge in data storage and brings with it all of the aforementioned risks in data interpretation.

These challenges illustrate the need for collaboration and synergistic activities in this field of research. There are several ways in which such efforts can be facilitated. From a laboratory perspective, information and details need to be widely distributed from the pioneers of new techniques to the people following in their footsteps. Likewise, groups offering low-resolution techniques can serve as the "find-r-scope" for those using high-resolution techniques, and collaboration must be fostered between these groups. Additionally, the lessons from microwave and infrared "stepping stone" techniques need to be shared with those doing the new technique development in the far-infrared.

# IV. Recommendations for Future Research Directions

New advances in far-IR laboratory spectroscopy are required to advance the field and meet the needs of new observatories such as ALMA and SOFIA, as well as ongoing needs from Herschel. This is the top priority for far-IR research efforts over the next few years. Other related research areas, including theory, modeling, collisional information, reaction dynamics, etc., are also very important, and efforts in these areas must have synergy with the lab efforts. Methanol will be used here as an example of the types of efforts that are needed.

## *1. Laboratory Spectroscopy*

Improvements and developments in laboratory spectroscopy are needed to support astronomical observations and models of interstellar chemistry. Two major areas of effort are anticipated. Because the spectroscopy of stable neutral molecules in their ground states is generally well known, future studies should focus on neutral radicals and molecular ions. Metastable excited states may also be of interest.

1) The production and identification of new radical and ion molecular spectra in the lab requires sources producing cold molecules, which improves the density-per-quantum state available, and high sensitivity detection schemes with the selectivity to identify new species. Because theory is not yet reliable enough to precisely estimate transition frequencies, spectral searches will require broad wavelength scans to search for new species whose spectra have not been detected before. Broad survey scans are usually done at relatively low resolution.

2) Low-resolution survey spectra will need to be refined at higher resolution in order to determine the unique spectral signatures required to make unambiguous identifications in astronomical spectra.

Topics 1) and 2) may require very different experimental approaches and specialized instrumentation not found in the same laboratories. The details of these experiments will be very different for electronic, vibrational or rotational spectra. Because new observational capabilities (ALMA, Herschel, SOFIA, and ultimately JWST) focus on lower frequencies, we focus the discussion here on new rotational and vibrational spectroscopy measurements.

### *1.1 Production of radicals and ions*

Radicals may be produced by thermal decomposition, ultraviolet photolysis or electrical discharges using suitable precursors. Ions are generally produced by discharges. The density of neutral radicals can be $10^{10}$-$10^{12}$/cm$^3$, while that for ions is generally much lower (~$10^6$/cm$^3$). In both cases, production of new species requires energetic conditions incompatible with high-resolution spectroscopy. Collisional cooling of some sort is required to achieve high densities in the lowest few quantum states. High pressure environments cause collisional line broadening and matrix isolation introduces spectral shifts that are difficult to predict.  Therefore, the ideal environment for both applications is supersonic molecular beams.

*1.2  Infrared Spectroscopy*

Vibrational frequencies for even small radicals and ions of astrophysical significance cannot presently be determined by theory with sufficient accuracy to enable high-resolution infrared spectroscopy to locate new spectra.  High-resolution laser scans over even tens of cm$^{-1}$ (i.e. hundreds of GHz) are impractical, and therefore new species are not generally found in this way.  Low-resolution (0.1-1.0 cm$^{-1}$ line width) laser scans covering broad ranges (600-4500 cm$^{-1}$) of the infrared are now available with optical parametric oscillator (OPO) systems.  In the past, these OPO's functioned only in the high-frequency region of C–H and O–H stretches, where species identification can be ambiguous.  Recent developments in technology have extended these measurements to the so-called infrared "fingerprint" region, where spectral signatures are more definitive.

Although cavity-enhanced IR absorption spectroscopy is possible in principle with OPO systems, the lack of high-reflectivity IR mirrors across the full spectrum has limited progress in this area.  New IR measurements of neutrals have more often involved double resonance IR/UV-visible methods, with fluorescence- or ionization-depletion involving known electronic transitions, where detection sensitivity is greater than that in IR absorption [Zwier 2001].  Progress has been somewhat greater for ion spectra using mass-selection and infrared photodissociation spectroscopy with the method of rare gas atom "tagging" [Yeh et al. 1989; Boo et al. 1995; Bieske & Dopfer 2000; Robertson & Johnson 2003].  Although predissociation introduces lifetime broadening in these spectra (line widths 1-10 cm$^{-1}$), and the rare gas atom induces some spectral shifts (<10 cm$^{-1}$), new spectra have been reported for small hydrocarbon cations over the full spectrum of the IR [Duncan 2012].  Introduction of double-resonance methods with the potential to eliminate the shifts and line width problems of tagging is a much-needed development.

High-resolution IR measurements suffer from limitations in scan range, as well as spectra that may overlap with precursors, but recent high-sensitivity measurements have been described.  Nesbitt has employed multipass IR absorption in slit jets, with discharge modulation to discriminate against precursors.  This method has adequate sensitivity for many neutrals [Davis et al. 2000; Dong & Nesbitt 2006a], and has had some success with ions [Dong & Nesbitt 2006b].  McCall has employed velocity modulation and cavity enhancement in the NICE-OHVMS method, and has had some initial success with ions [Crabtree et al. 2006].  Both methods have limited search capability and require input for band positions from low-resolution survey scans.

*1.3 Microwave/Terahertz Spectroscopy*

Efficient detection of neutral radicals and ions requires the continued development of fast and sensitive spectrometers that can be easily coupled to production sources. These spectrometers should ideally operate in the THz/submillimeter range to provide the best spectral overlap with the capabilities of ALMA, Herschel HIFI, and SOFIA. However, spectra in these frequency ranges can be quite congested and are seldom interpretable without data at microwave/centimeter frequencies, particularly for species containing more than a few non-hydrogen atoms.

In the microwave, all of the necessary pieces appear to be in place. The chirped-pulse method can be used to rapidly acquire survey scans, and chirped-pulse molecular beam spectrometers have been successfully coupled to pulsed discharge sources [Karunatilaka et al. 2010]. These same sources have also been coupled to Balle-Flygare-type spectrometers that use cavity enhancement factors to obtain extremely high resolutions and sensitivities over narrow bandwidths in a small amount of lab time [McCarthy & Thaddeus 2010]. In this frequency range, the bottleneck appears to be the total number of spectrometers worldwide that combine these capabilities rather than any particular limitations of the capabilities themselves.

The technology that has enabled THz spectroscopy is still maturing. Sources of THz radiation have been steadily improving in output power, tunability, and frequency coverage, and these improvements have coincided with the development of spectroscopic techniques that achieve significantly higher sensitivity per unit time than conventional direct absorption measurements. The FAst Scan Submillimeter Spectroscopic Technique (FASSST) [Petkie et al. 1997; Medvedev et al. 2004] developed by De Lucia and co-workers can rapidly collect high-resolution survey spectra, and there have also been promising recent extensions of chirped-pulse techniques [Gerecht et al. 2011; Steber et al. 2012] into these frequency ranges. The primary needs here are improved sources and detectors/detection schemes (such as cavity ringdown techniques [Meshkov et al. 2005]) to further enhance instrument sensitivity. Work also needs to be done to efficiently couple these instruments with neutral radical and ion production sources.

*2. Theoretical Advances in Computational Spectroscopy*

Research in astrochemistry relies on reliable, calculated line-list spectra of interstellar molecules, as well as accurate collision cross sections for radiative transfer. Calculated spectral line-lists provide an essential complement to experimentally measured ones. One obvious way that theory can complement experiment is in "filling in the gaps" in laboratory-measurements. Theoretical line-lists can in principle span the entire spectral range. Another major computational challenge in astrochemistry is determination of accurate collision cross sections for radiative transfer, as many observations probe regions that are not in local thermodynamic equilibrium. The goals for these two areas of theoretical/computational research have many similar points, which are overviewed below and in the next section. Here we continue to use the example molecule methanol that is of wide interest and also challenging to theory. Methanol is ubiquitous in the ISM, and has also been the focus of numerous experiments. Below is a list of what is needed to advance computational spectroscopy and comments on the challenges, in particular for $CH_3OH$.

- Accurate ab initio potentials. For $CH_3OH$ this potential is a function of 12 degrees of freedom.
- Accurate ab initio dipole moment surfaces. For $CH_3OH$ each component is a function of 12 degrees of freedom.
- Development of essentially quantum methods to calculate rovibrational wavefunctions. For $CH_3OH$ this requires an accurate treatment of the facile internal torsional mode.
- The calculated line list, and corresponding spectral simulations as a function of temperature. For $CH_3OH$ the line list consists of millions of lines which must be cataloged and made available to observers.
- Refinement and testing of "spectroscopic Hamiltonians" and development of line-list extensions.

The state-of-the-art is *almost* at the level where the above program can be carried out for $CH_3OH$ [Xu and Hougen 1995; Xu 2000; Fehrensen et al. 2003; Miani et al. 2000; Hanninen and Halonen 2003; Sibert and Castillo-Chara 2005; Bowman et al. 2007; Carter et al. 2009] and other molecules of astrochemical interest, including a variety of carbocations and $H_5^+$. $H_5^+$ in particular is an example of a highly fluxional system. Such systems present additional challenges to computational methods, because, unlike $CH_3OH$, which has one large amplitude mode, fluxional systems have more than one such mode. Progress is being made in this area, in particular for $H_5^+$, and a number of papers have appeared in the literature presenting calculations that approach essentially exact results [Cheng et al. 2012; Valdés et al. 2012].

Ab initio electronic structure codes, e.g., MOLPRO [Werner et al. 2008] and MOLCAS [Aquilante et al. 2010], now can be used routinely to perform very high level electronic structure calculations of energies and dipole moments. These can be represented by force fields or more global potential energy surfaces and special expressions for the dipole moment fitting [Braams & Bowman 2009]. Several codes, notably MULTIMODE, can be used to obtain virtually exact line-list spectra for tetraatomic and larger molecules [Carter & Bowman 2011; Carter & Bowman 2012].

While such computational studies are now feasible for smaller molecular species, extensions to larger molecular systems remain a computational challenge. Currently, full spectral treatment for molecules with >6 atoms is limited to the use of effective Hamiltonians, which often do not enable extrapolation beyond laboratory measurements. Even these Hamiltonians are incomplete for fluxional species, where often a Hamiltonian that accurately describes one fluxional system cannot be extended to other systems.

Accurate calculations of collision cross sections (and from them the rate coefficients) of radiative transfer also require the existence of ab initio potential energy surfaces. Unlike the potentials for spectroscopy, the potentials for collisional processes must be global, that is they must describe the collision partners as well as the strong interaction region. Software such as MOLSCAT [Hutson & Green 1994] and Hibridon [Alexander et al. 2011] are available for inelastic scattering, and so the major effort for future studies is in developing the global potentials. For methanol collisions with $H_2$, a full-dimensional surface has 18 degrees of freedom, though if the computation is limited to rotational excitation, a 5D surface can be adopted [Rabli & Flower 2012].

## 3. Astrochemical Modeling

In addition to the important need for laboratory and theoretical spectroscopy of complex organic molecules (COMs) in the FIR/submm to identify observed spectral features, interpretation of line intensities depends critically on a range of laboratory data. Modeling line intensities in low-density environments requires predictions of the COM abundances and, for the case of emission features, excited state populations. While at high density, both level populations and species abundances can be accurately predicted in LTE, the density of most interstellar environments is sufficiently low that they must be treated in non-LTE, and typically in a time-dependent fashion. This typically involves a network of rate equations for the important processes. Overall, the predominant need for future modeling studies are support for astrochemical sensitivity studies to determine which gas-phase, gas-grain, and inelastic excitation processes require detailed laboratory and theoretical investigation. The current state-of-the-art approaches and challenges in each of these areas are overviewed below.

### 3.1. Chemistry

Reaction networks for astrochemistry have developed over many decades and initially included only gas-phase processes. Later it was recognized that the formation of some molecules, including some COMs, via gas-phase reactions was not sufficient to explain observations, and that grain-catalyzed processes were important (see Tielens [2005] and references therein). While the basic mechanisms for both gas-phase and gas-grain chemistry are understood, what is generally lacking is quantitative information on the fundamental input parameters (rate coefficients, diffusion rates, activation barriers, etc.). Astrochemical network databases (e.g., KIDA, Wakelam et al. [2012]) include ~4000 gas-phase reactions for 100s of species. While there has been progress in both experimental and theoretical methods, ~80% of the reaction rate coefficients in such networks are guesstimates. For reactions that have been studied in detail, the results are typically limited in scope.

The key needs for future gas-phase reaction studies include extension to higher temperature (1000 K), resolution of para- and ortho-moieties, resolution of final-states, consideration of initial excited states, and investigation of ionization processes.

The workhorses for computational reactive scattering studies are the quasiclassical trajectory (QCT) method [Hase 1988] and the quantum wave-packet propagation (QWP) approach (see Althorpe and Clary [2003] for a review). In principle, while less accurate, QCT studies can be performed on any size collision system, the state-of-the art in QWP calculations is H+CH$_4$. Both approaches depend on the accuracy and availability of potential energy surfaces (PES) and, for open-shell or ion-collision systems, surfaces of nonadiabatic couplings. Currently, full-dimensional PESs for systems up to 10 atoms can be computed and fitted [Braams & Bowman 2009], but dynamics on such large systems are limited to QCT studies. Today, it is not possible to perform quantum reactive studies involving methanol except with a reduced-dimensional approach.

Key needs include support for i) development of QWP methods for COMs, ii) computational studies of COMs by QCT, and iii) generation of COM relevant PESs. Further, support for development of new experimental techniques for measurement of COM reactions to benchmark calculations will be needed.

For gas-grain chemistry (e.g. Garrod et al. [2008]), the networks are almost exclusively based on empirical estimates of fundamental parameters (i.e., sticking coefficients, molecule-surface binding energies and vibrational frequencies, activation energies, surface barriers, etc.). Even for the sticking of atomic H on amorphous water ice, significant discrepancies exist between theory, experiment, and simulation. Our knowledge of surface interactions for more complicated radicals (e.g., CH$_2$OH, CH$_3$O) is far more uncertain.

The key needs for future modeling studies include support for laboratory measurements of fundamental parameters and simulational studies of gas-grain interaction parameters, and the effects of non-thermal desorption, photodesorption, and X-ray desorption.

*3.2. Non-LTE Spectral Emission Modeling*

The intensity of emission lines is directly related to the population of the emitting level, which in turn is controlled by a competition of radiative and collisional processes. In many environments, the gas density is below the critical density for a given level, such that collisional excitation rate coefficients are required to model the emission. Van der Tak [2011] recently reviewed the status of available collisional data relevant to FIR/submm observations.

The primary workhorse for inelastic collisional studies is the time-independent close-coupling (CC) method which, similar to reactive scattering, requires accurate PESs. The largest systems for which full-dimensional CC calculations have been performed are diatom-diatom (i.e., 4 atoms) collisions, though reduced-dimensional (rigid-rotor) calculations have been performed for low-lying rotational levels for systems as large as $HCOOCH_3$-He. While software, as mentioned above, is available for the scattering calculations, such methods are limited to treatment of rotational excitation, while for vibrational excitation only atom-diatom systems can be studied. Very little methodical development has occurred for the past three decades and few benchmarking experiments are available.

The key needs in non-LTE spectral emission modeling are support for full-dimensional inelastic scattering method development, full-dimensional PES calculations, experimental method development, treatment of bending, torsional, and umbrella excitation modes, development of approximate methods to estimate COM collisional cross-sections in a quasi-classical fashion, and ultimately extension of all approaches to COMs.

## V. Summary and Conclusions

The participants of the workshop overviewed the state-of-the art in molecular spectroscopy in support of far-infrared astronomy, and identified the key needs to advance this field of research over the next 5-10 years. The development of new laboratory capabilities that offer rapid, broadband, high-resolution, high-sensitivity measurements of molecular spectra in the frequency ranges that overlap with new observational facilities was identified as the top priority for future far-infrared laboratory astrophysics initiatives. These laboratory efforts must be supported by additional efforts in theory, computation, and modeling that push beyond current techniques to study larger molecular systems. Additionally, development of suitable analysis tools for broadband molecular spectra from both the laboratory and space is critical.

**Appendix 1. Status of Efforts to Characterize the Interstellar "Weeds" (contributed by John Pearson)**

A workshop on Laboratory astrophysics in support of the Herschel Space observatory was held in October 2006 at the California Institute of Technology in Pasadena California. The final report detailed the importance of performing laboratory spectroscopy at frequency commensurate with Herschel, i.e. up to 1.9 THz for the abundant molecules with rich spectra. The most important four molecules (Class 1 "weeds") identified are $CH_3OH$ (methanol), $HCOOCH_3$ (methyl formate), $CH_3OCH_3$ (dimethyl ether), $CH_3CH_2CN$ (ethyl cyanide), and their isotopologues. Five somewhat less prominent molecules (Class 2 "weeds") are $C_2H_3CN$ (vinyl cyanide), $SO_2$ (sulfur dioxide), $CH_3CN$ (methyl cyanide), $HC_3N$ (cyanoacetylene), and $CH_3CHO$ (acetaldehyde).

Subsequently there was a Herschel call for laboratory astrophysics along with several rounds of the annual laboratory portion of the NASA APRA program. Similar opportunities also occurred in France, Germany, Canada and Spain. A number of proposals were funded and some rather limited funding to extend frequency laboratory frequency coverage was made available. The spectroscopic community made unprecedented efforts to collaborate and address the weeds for Herschel. The result of this effort is one of the great successes of the Herschel program and an enormous help for ALMA and SOFIA.

*A. 1. The status of the Class 1 weeds today*

Methanol (NASA APRA, Canada, Germany, France)

$^{12}CH_3^{16}OH$: The main isotopologue of methanol has been completed to the level needed for Herschel. There are a number of known issues with the large quantum number wave functions, but currently funded efforts in New Brunswick Canada and JPL are sorting these out (very few of these lines are in Herschel HIFI spectra). This should lead to a good understanding the $v_t$=0, 1, 2 states to J>40 and K=19, 17, 15, respectively. A very preliminary check of the intensities using HIFI calibration funds was performed at JPL involving a limited sampling of torsional transitions. This study observed significant scatter, but conclude that on average the intensity model including the three-fold term was a reasonable approximation of reality. Subsequent microwave studies have shown some more systematic intensity deviation in frequency ranges where ground based observations are commonly made. Methanol appears to have been anointed as the "Swiss army knife" of astrophysics due to just about any small frequency windows containing lines covering a wide range of energies. A comprehensive study of intensities in any internal rotor has never been performed, and considering the way methanol will be used astronomically, this effort is a particularly important problem astrophysics and would be of great value to SOFIA and in subsequent investigations into Herschel data. ALMA and JWST may be able to observe the excited vibrational states that have thus far defied a theoretical description. If these instruments do see these states, then sorting out the molecular physics of the previously unsolved problem of coupled large amplitude motion-normal vibration could become a priority. Theoretical calculations do have the potential to assist greatly in this effort. None of this will be easy or inexpensive and should not be a high priority until there is an observational need for the work.

$^{13}CH_3^{16}OH$: Two studies from Cologne Germany with New Brunswick Canada and Toyama Japan are nearing publication. Extensive data from JPL has been collected extending the J and K coverage of these studies, and analysis will be performed the other datasets are published. This is certainly sufficient for Herschel and SOFIA, but ALMA may require measurements of more torsional states and quantum numbers.

$^{12}CH_3^{18}OH$: Extensive data has been collected, and assignments will focus on the ground state. Existing data has been good enough for Herschel spectral interpretation.

$^{12}CH_3^{16}OD$: A study form Cologne, Germany, and New Brunswick Canada is nearing completion. Additional data has been collected at JPL should the quantum numbers need to be extended. Good catalog files are the top priority, but these should come out of the present efforts.

$CH_2DOH$: A JPL study of the ground state (funded by a Herschel–related laboratory astrophysics grant) has been published. A joint French (infrared) microwave study of the more excited states is in progress. The molecular physics is still at the effective Hamiltonian stage. Improved intensities are still necessary.

$CHD_2OH$: No significant work has been done. This will be a problem for cold core studies with ALMA. The lines will be below the noise level with Herschel and SOFIA observations.

$CD_3OH$. The data is sufficient for cold core studies where only the ground state is populated.

Methyl Formate (NASA APRA, French)

All the major isotopes of methyl formate have been analyzed in the lowest two torsional states. Above $v_t=1$ there appears to be some issues with higher lying small amplitude vibrational states. An enormous amount of data has been collected and analysis is on-going in Russia and France. The data appears to be adequate for Herschel and SOFIA, however ALMA is likely to see more excited states.

Dimethyl Either (NASA APRA, German)

$^{12}CH_3O^{12}CH_3$: The ground state has been studied through a JPL/Cologne/University of Missouri Kansas City collaboration and completed to the level necessary for Herschel. The two lowest torsional states have also been studied and are in the process of being written up for publication.

$^{13}CH_3O^{12}CH_3$: The ground state has been studied by a collaboration between Cologne and the University of Missouri Kansas City to the level necessary for Herschel.

$CH_2DOCH_3$: Has been investigated to the level required for detection.

Ethyl Cyanide (NASA APRA, NASA Herschel, French)

$^{12}CH_3^{12}CH_2^{12}C^{14}N$: The ground state has been done to J=130 and is solved for astrophysics (NASA APRA result). The two lowest vibrational states (the torsion and the in-plane CCN bend) are known and

nearing publication. A partial analysis of the CCN out of plane bend (the third lowest vibrational state) and the in-plane CCC bend (the 7th lowest vibrational state) are in press. Limited assignments in the states between have proven that all can be identified in ground base spectra before ALMA. The physics of coupled large amplitude motion-normal vibration remains an unsolved physics problem and is the central challenge in ethyl cyanide. A complete quantum mechanical analysis will require years of effort and is an example of where other studies may be mandatory in the short term. A nearly complete spectrum exists along with some temperature dependent spectra, collected at Ohio State.

$^{13}$C ethyl cyanide: Ground states assignments are finished. Intensities suggest that the coupled bend and torsion problem needs to be solved for all. Present APRA effort is funding a joint JPL/Lille effort to complete this analysis.

$^{14}$N ethyl cyanide: Ground state assignments are finished. ALMA might detect higher vibrational states, but this work is not as much of a priority as analyzing the vibrations in the main isotopologue.

Deuterated ethyl cyanide: Several versions have been studied at some level, but most of these lines are expected to be in the noise for most observational studies. Double $^{13}$C varients are probably more important, but remain to be done.

Status of Class 2 Weeds Today

C$_2$H$_3$CN (vinyl cyanide) (NASA APRA, Poland).

$^{12}$C$_2$H$_3$$^{12}$C$^{14}$N: The first five vibrational states are known to high J and K values. Data is sufficient for single dish observations but is likely not to include enough vibrational states for ALMA
$^{13}$C$_2$H$_3$$^{12}$C$^{14}$N: Done sufficiently for SOFIA and Herschel, more might be needed for ALMA
$^{12}$C$_2$H$_3$$^{13}$C$^{14}$N: Done sufficiently for SOFIA and Herschel, more might be needed for ALMA
$^{12}$C$_2$H$_3$$^{12}$C$^{15}$N: Done sufficiently for SOFIA and Herschel, more might be needed for ALMA

SO$_2$ (sulfur dioxide),
The spectroscopy of SO$_2$ was believed to be sufficient for Herschel and no additional work was performed. ALMA may need more coverage of vibrational states.

CH$_3$CN (methyl cyanide), (NASA APRA work, Germany, France)

$^{12}$CH$_3$$^{12}$C$^{14}$N The ground and lowest vibrational state have been measured to high J values (Cologne/JPL). Data on many more excited vibrational states is in preparation (Cologne/JPL/France).
$^{13}$C/$^{14}$N: The ground state has been done to high J (Cologne/JPL).
Double $^{13}$C: The ground state has been done to high J (Cologne/JPL).
CH$_2$DCN: The ground state has been done to high J (Cologne/JPL).
ALMA might need the $v_8$=1 state of the more abundant isotopic species. Lab data exists, so only the analysis would be needed.

HC$_3$N (cyanoacetylene),

Our knowledge of spectra of this linear molecule has proven to be good enough for Herschel. More vibrational states for the isotopologues will be important for ALMA.

CH$_3$CHO (acetaldehyde) (NASA Herschel, NASA APRA, France)

The ground state is done to modestly high J ~60 (NASA Herschel Lab-Astrophysics task). Torsional states are done to the point where they begin to interact with the v$_{10}$ vibration. An attempt was made (NIST/Paris) to address the interactions of the torsion coupled small amplitude vibration. It was the best analysis to date, but was not satisfactory to the level required by radio astronomy. Data has proven to be adequate for Herschel. The problems begin at 700K, so ALMA will require more work to be done. There is no shortage of laboratory spectra. Analysis is the challenge. Several isotopic species have also been done, including CH$_2$DCHO. More analysis will be forthcoming France/Russia/JPL. Data is sufficient for NASA but probably not ALMA.

Summary:

The weeds have been largely dealt with to the level required for single dish observations such as those of Herschel and SOFIA. The one notable exception is an exploration of the intensities in the three fold internal rotation case where theory has never been experimentally tested. The recommendation is to follow astronomy and perform the benchmark studies on the main isotopologue of methanol. ALMA needs will determine how spectroscopists need to go in solving the problem of coupled large and small amplitude motions. This problem becomes central in the analysis of many of these species in their next few excited vibrational and torsional states. For NASA, any major effort should be deferred until there is a well-established observational need. The NSF has already identified the problem in ALMA science verification datasets.